\documentclass[reqno,11pt]{amsart}
\usepackage[utf8]{inputenc}
\usepackage{graphicx}
\usepackage{amscd}
\usepackage{slashed}
\usepackage{amssymb}
\usepackage{esint}
\usepackage{enumitem}
\usepackage[off]{auto-pst-pdf} 
\usepackage{pst-grad} 
\usepackage{pst-plot} 
\usepackage[mathscr]{eucal}
\numberwithin{equation}{section}
\allowdisplaybreaks[1]

\newcommand{\SetFigFont}[3]{}

\title[The Fermionic Signature Operator and Space-Time Symmetries]{The Fermionic Signature Operator \\ and
Space-Time Symmetries}

\author[F.\ Finster]{Felix Finster}
\address{Fakult\"at f\"ur Mathematik \\ Universit\"at Regensburg \\ D-93040 Regensburg \\ Germany}
\email{finster@ur.de}
\author[M.\ Reintjes]{Moritz Reintjes \\ \\ August 2017}
\address{Departamento de Matem{\'a}tica \\ Instituto Superior T{\'e}cnico \\ 1049-001 Lisbon \\ Portugal}
\email{moritzreintjes@gmail.com}
\thanks{M.R.\ is partially supported by FCT/Portugal through (GPSEinstein) PTDC/MAT-ANA/1275/2014 and UID/MAT/04459/2013.}

\newtheorem{Def}{Definition}[section]
\newtheorem{Thm}[Def]{Theorem}
\newtheorem{Prp}[Def]{Proposition}
\newtheorem{Lemma}[Def]{Lemma}

\newtheorem{Corollary}[Def]{Corollary}
\newtheorem{Example}[Def]{Example}

\newcommand{\Thanks}{\vspace*{.5em} \noindent \thanks}
\newcommand{\beq}{\begin{equation}}
\newcommand{\eeq}{\end{equation}}
\newcommand{\Proof}{\begin{proof}}
\newcommand{\QED}{\end{proof} \noindent}
\newcommand{\QEDrem}{\ \hfill $\Diamond$}

\newcommand{\bra}{\mathopen{<}}
\newcommand{\ket}{\mathclose{>}}
\newcommand{\Sl}{\mathopen{\prec}}
\newcommand{\Sr}{\mathclose{\succ}}

\newcommand{\C}{\mathbb{C}}
\newcommand{\R}{\mathbb{R}}
\newcommand{\1}{\mbox{\rm 1 \hspace{-1.05 em} 1}}

\renewcommand{\H}{\mathscr{H}}

\newcommand{\G}{\mathcal{G}}
\newcommand{\U}{{\mathcal{U}}}

\newcommand{\bep}{\begin{pmatrix}}
\newcommand{\enp}{\end{pmatrix}}

\renewcommand{\O}{\mathscr{O}}

\newcommand{\Dir}{{\mathcal{D}}}
\newcommand{\D}{{\mathscr{D}}}

\newcommand{\B}{{\mathscr{B}}}
\renewcommand{\O}{{\mathscr{O}}}

\newcommand{\Lin}{\text{\rm{L}}}
\newcommand{\Cisc}{C^\infty_{\text{\rm{sc}}}}
\newcommand{\Cisco}{C^\infty_{\text{\rm{sc}},0}}

\DeclareMathOperator{\supp}{supp}

\newcommand{\p}{\mathfrak{p}}
\newcommand{\Sig}{\mathscr{S}}
\newcommand{\scrM}{\mycal M}
\newcommand{\scrN}{\mycal N}
\newcommand{\g}{{\rm{g}}}
\newcommand{\h}{{\rm{h}}}
\newcommand{\e}{{\rm{e}}}

\DeclareFontFamily{OT1}{rsfso}{}
\DeclareFontShape{OT1}{rsfso}{m}{n}{ <-7> rsfso5 <7-10> rsfso7 <10-> rsfso10}{}
\DeclareMathAlphabet{\mycal}{OT1}{rsfso}{m}{n}

\begin{document}

\maketitle
\begin{abstract}
We show that and specify how space-time symmetries give
rise to corresponding symmetries of the fermionic signature operator
and generalized fermionic projector states.
\end{abstract}

\tableofcontents

\section{Introduction}
The fermionic signature operator introduced in~\cite{finite, infinite}
provides a setting of spectral geometry in Lorentzian signature~\cite{drum}
and has been proven useful for constructing quasi-free Dirac states
in globally hyperbolic space-times~\cite{fewster+lang, hadamard}.
In all known examples, the resulting so-called fermionic projector state
respects the symmetries of space-time.
The present paper is devoted to a systematic study of the general relationship
between space-time symmetries
and symmetries of the fermionic signature operator.
We also study the symmetry properties of the resulting generalized fermionic projector states.

We describe space-time symmetries by a Lie group~$\G$ which
acts locally as isomorphisms~$\Phi$ of the spinor bundle~$S\scrM$
(see Section~\ref{secsymmst}). Considering such local actions has the
advantage that they can be obtained from Killing symmetries in a
straightforward manner (see Section~\ref{seckilling}).
The only condition needed is that the Killing fields are complete (see Definition~\ref{defunifflow}).
We construct a strongly continuous action of~$\G$ represented
by unitary operators on the Hilbert space of Dirac solutions
(Theorem~\ref{thmstrong}).
The corresponding Lie symmetries are represented by essentially self-adjoint
operators acting on the smooth and spatially compact solutions
(Theorem~\ref{thmalgebra}).
Both the local representation of the group and the representation of the Lie algebra
on the solution space commute with the fermionic signature operator (see Theorems~\ref{thmSfinite},
\ref{thmSinfinite} and~\ref{thmalgebra}), except for a minus sign which appears
if the time orientation is reversed.
Moreover, the resulting generalized fermionic projector states are
invariant under the symmetry transformations, again up to signs
(Theorem~\ref{thmgenFP} and Corollary~\ref{corinf}).
As applications we consider Killing symmetries and discrete symmetries
(Section~\ref{secappl}).
The paper concludes with a discussion of several examples (Section~\ref{secex}).

\section{Preliminaries}
\subsection{Lorentzian Spin Geometry}
Let~$(\scrM, g)$ be a smooth, globally hyperbolic, time-oriented Lorentzian spin
manifold of dimension~$k \geq 2$.
For the signature of the metric we use the convention~$(+ ,-, \ldots, -)$.
We denote the corresponding spinor bundle by~$S\scrM$. Its fibers~$S_p\scrM$ are endowed
with an inner product~$\Sl .|. \Sr_p$ of signature~$(n,n)$
with~$n=2^{[k/2]-1}$ (where~$[.]$ is the Gau{\ss} bracket; for details see~\cite{baum, lawson+michelsohn}),
which we refer to as the spin scalar product.
Clifford multiplication is described by a mapping~$\gamma$
which satisfies the anti-commutation relations,
\[ \gamma \::\: T_p\scrM \rightarrow \Lin(S_p\scrM) \qquad
\text{with} \qquad \gamma(u) \,\gamma(v) + \gamma(v) \,\gamma(u) = 2 \, g(u,v)\,\1_{S_p(\scrM)} \:. \]
We also write Clifford multiplication in components with the Dirac matrices~$\gamma^j$.
The metric connections on the tangent bundle and the spinor bundle are denoted by~$\nabla$.
The sections of the spinor bundle are also referred to as wave functions.

In order to include the situation when an external 
potential is present, we introduce a multiplication operator~$\B(p) \in \Lin(S_p\scrM)$, which we assume to
be smooth and symmetric with respect to the spin scalar product,
\beq \label{Bdef}
\B \in C^\infty(\scrM, \Lin(S\scrM)) \qquad \text{with} \qquad
\Sl \B \phi | \psi \Sr_p = \Sl \phi | \B \psi \Sr_p \quad \forall \phi, \psi \in S_p\scrM\:.
\eeq

\subsection{The Dirac Operator and Inner Products on Wave Functions}
We denote the smooth sections of the spinor bundle by~$C^\infty(\scrM, S\scrM)$.
Similarly, $C^\infty_0(\scrM, S\scrM)$ are the smooth sections with compact support.
On the wave functions, one has the Lorentz invariant inner product
\begin{gather*}
\bra .|. \ket \::\: C^\infty(\scrM, S\scrM) \times C^\infty_0(\scrM, S\scrM) \rightarrow \C \:, \notag \\
\bra \psi|\phi \ket = \int_\scrM \Sl \psi | \phi \Sr_p \: d\mu_\scrM\:.
\end{gather*}
The Dirac operator~$\Dir$ is defined by
\beq \label{Dirdef}
\Dir := i \gamma^j \nabla_j +\B \::\: C^\infty(\scrM, S\scrM) \rightarrow C^\infty(\scrM, S\scrM)\:.
\eeq
For a given real parameter~$m \in \R$ (the ``mass''), the Dirac equation reads
\[ (\Dir - m) \,\psi_m = 0 \:. \]
For clarity, we always denote solutions of the Dirac equation by a subscript~$m$.
We mainly consider solutions in the class~$\Cisc(\scrM, S\scrM)$ of smooth sections
with spatially compact support. On such solutions, one has the scalar product
\beq \label{print}
(\psi_m | \phi_m)_m = 2 \pi \int_\scrN \Sl \psi_m \,|\, \gamma(\nu)\, \phi_m \Sr_p\: d\mu_\scrN(p) \:,
\eeq
where~$\scrN$ denotes any Cauchy surface and~$\nu$ its future-directed normal
(due to current conservation, the scalar product is
in fact independent of the choice of~$\scrN$; for details see~\cite[Section~2]{finite}).
Forming the completion gives the Hilbert space~$(\H_m, (.|.)_m)$.
It will be convenient to use the short notation
\[ \H_m^\infty := \H_m \cap \Cisc(\scrM, S\scrM) \:. \]

\subsection{The Fermionic Signature Operator in Finite Lifetime}
We now recall the construction of the fermionic signature operator in~\cite{finite},
which applies in particular to space-times of finite life-time.
We here consider a slightly more general setting which also
applies to certain space-times involving horizons like the Rindler space-time~\cite{rindler}.

\begin{Def} \label{defmfinite}
The manifold~$(\scrM,g)$ is said to be {\bf{weakly {\em{m}}-finite}} if
for every~$\phi_m \in \H_m^\infty$, there is a
constant~$c(\phi_m)>0$ such that for all~$\psi_m \in \H_m^\infty$, the
function~$\Sl \psi_m | \phi_m \Sr_p$  is integrable on~$\scrM$ and
\beq \label{weakbound}
|\bra \psi_m | \phi_m \ket| \leq c\:\|\psi_m\|_m\:.
\eeq
\end{Def}
Under this assumption, the Fr{\'e}chet-Riesz theorem gives rise to a unique densely defined operator
\[ \Sig_m \::\: \H_m^\infty \rightarrow \H_m \]
with the property
\[ \bra \psi_m | \phi_m \ket = (\psi_m \:|\: \Sig_m\, \phi_m)_m \qquad \text{for all ~$\psi_m \in \H_m$}\:, \]
referred to as the {\bf{fermionic signature operator}}.
We remark that the notion of {\bf{{\em{m}}-finiteness}} in~\cite{finite} instead of~\eqref{weakbound} imposes the stronger
assumption
\beq \label{stbound}
|\bra \psi_m | \phi_m \ket| \leq c\: \|\phi_m\|_m \:\|\psi_m\|_m\:.
\eeq
This inequality holds in every space-time of finite lifetime.
It implies that the fermionic signature operator is bounded and can thus be extended
to all of~$\H_m$.

\subsection{The Fermionic Signature Operator in Infinite Lifetime}
In a space-time of infinite life time~\cite{infinite}, one studies families of solutions.
More precisely, we consider the mass parameter in a bounded open interval, $m \in I := (m_L, m_R)$
with~$0 \not\in I$. By~$\Cisco(\scrM \times I, S\scrM)$ we
denote the smooth wave functions with spatially compact support which
are also compactly supported in~$I$. 
We often denote the dependence on~$m$ by a subscript, $\psi_m(p) := \psi(p,m)$.
On families of solutions in~$\Cisco(\scrM \times I, S\scrM)$,
for any fixed~$m$ we can take the scalar product~\eqref{print}.
We introduce a scalar product on families of solutions by integrating over the mass parameter,
\[ ( \psi | \phi) := \int_I (\psi_m | \phi_m)_m \: dm \]
(where~$dm$ is the Lebesgue measure). Forming the completion gives the
Hilbert space~$(\H, (.|.))$. We denote the norm on~$\H$
by~$\| . \|$. Moreover, we set
\[ \H^\infty := \H \cap \Cisco(\scrM \times I, S\scrM) \:. \]

We introduce~$T$ as the operator of multiplication by the mass parameter,
\[ T \::\: \H \rightarrow \H \:,\qquad (T \psi)_m = m \,\psi_m \:. \]
Integrating over~$m$ gives the operation
\[ \p \::\: \H^\infty \rightarrow \Cisc(\scrM, S\scrM)\:,\qquad
\p \psi = \int_I \psi_m\: dm \:. \]

\begin{Def} \label{defweakMOP}
The Dirac operator~$\Dir$ has the {\bf{weak mass
oscillation property}} in the interval~$I \subset \R$ with domain~$\H^\infty$
if the following conditions hold:
\begin{itemize}[leftmargin=2em]
\item[\rm{(a)}] For every~$\psi, \phi \in \H^\infty$, the
function~$\Sl \p \phi | \p \psi \Sr$ is integrable on~$\scrM$.
Moreover, for any~$\psi \in \H^\infty$ there is a constant~$c(\psi)$ such that
\[ |\bra \p \psi | \p \phi \ket| \leq c\, \|\phi\| \qquad \forall\: \phi \in \H^\infty \:. \]
\item[\rm{(b)}] For all~$\psi, \phi \in \H^\infty$,
\[ \bra \p T \psi | \p \phi \ket = \bra \p \psi | \p T \phi \ket \:. \]
\end{itemize}
\end{Def}
This definition specifies
the minimal requirements needed for the construction of the fermionic projector.
More precisely, under these assumptions, the Fr{\'e}chet-Riesz theorem
gives rise to a densely defined symmetric operator~$\Sig$ acting on families of solutions defined by
\beq \label{Sigdef}
\Sig \::\: \H^\infty \rightarrow \H \:,\qquad
(\Sig \psi | \phi) = \bra \p \psi | \p \phi \ket \quad \forall\: \phi \in \H\:.
\eeq
This operator is shown to commute with~$T$.
After constructing the Friedrichs extension of~$\Sig^2$, the spectral theorem
for commuting operators gives rise to a joint spectral measure~$dE_{\rho, m}$ of
the commuting operators~$\Sig^2$ and~$T$.
For the technical details of these constructions we refer to~\cite[Section~3]{infinite}.
This spectral measure makes it possible to define
the fermionic signature operator~$d\Sig_m$ as an operator-valued measure by
\beq \label{dSdef}
\int_I \eta(m)\: d\Sig_m := \int_{\sigma(\Sig^2) \times I} \eta(m)\: \Sig \: dE_{\rho,m}
\eeq
(for any test function~$\eta \in C^0(I)$). This formula also makes it possible to introduce a
functional calculus for~$\Sig_m$ in a straightforward way by
\begin{align*}
&\int_I \eta(m)\: dW(\Sig_m) \notag \\
&\;\;:= \int_{\sigma(\Sig^2) \times I} \eta(m)
\Big[ W \big( \sqrt{\rho} \big) \:\big( \Sig + \sqrt{\rho} \big)
+ W\big( -\sqrt{\rho} \big) \:\big( -\Sig + \sqrt{\rho} \big) \Big] \: \frac{dE_{\rho,m}}{2 \sqrt{\rho}}\:,
\end{align*}
where~$W$ is a bounded Borel function.

In order to construct the fermionic signature operator~$\Sig_m$
pointwise for any~$m \in I$, one needs a stronger assumption:
\begin{Def} \label{defstrongMOP}
The Dirac operator~$\Dir$ has the {\bf{strong mass oscillation property}}
in the interval~$I=(m_L, m_R)$ with domain~$\H^\infty$ if there is a constant~$c>0$ such that
\[ |\bra \p \psi | \p \phi \ket| \leq c \int_I \, \|\phi_m\|_m\, \|\psi_m\|_m\: dm
\qquad  \textnormal{for all} \: \psi, \phi \in \H^\infty\:. \]
\end{Def}

\noindent The following theorem is proved in~\cite[Theorem~4.2, Proposition~4.3 and
Theorem~4.7]{infinite}:
\begin{Thm} \label{thmSrep}
Assume that the Dirac operator~$\Dir$ has the strong mass oscillation property in the
interval~$I=(m_L, m_R)$. Then there exists a family of self-adjoint linear operators $(\Sig_m)_{m \in I}$
with~$\Sig_m \in \Lin(\H_m)$ which are uniformly bounded,
\[ \sup_{m \in I} \| \Sig_m\| < \infty\:, \]
such that
\beq \label{Spoint}
\bra \p \psi | \p \phi \ket = \int_I (\psi_m \,|\, \Sig_m \,\phi_m)_m\: dm \qquad
\text{for all~$\psi, \phi \in \H^\infty$}\:.
\eeq
The operator~$\Sig_m$ is uniquely determined for every~$m \in I$ by demanding that
for all~$\psi, \phi \in \H^\infty$, the functions~$( \psi_m | \Sig_m \phi_m)_m$ are continuous in~$m$.
Moreover, the operator~$\Sig_m$ is the same for all choices of~$I$ containing~$m$.
\end{Thm} \noindent

\subsection{Quasi-Free Dirac Fields and Generalized Fermionic Projector States}
We now explain the connection to quantum field theory
as worked out in~\cite{hadamard}.
Assume that the fermionic signature operator is bounded
(as is the case if space-time is strongly $m$-finite or if
the strong mass oscillation property holds).
Then the {\em{fermionic projector}}~$P$ is introduced as the operator
(for details see~\cite[Section~3]{finite} and~\cite[Section~4.2]{infinite})
\beq \label{Pstar}
P = -\chi_{(-\infty, 0)}(\Sig_m)\, k_m \::\: C^\infty_0(\scrM, S\scrM) \rightarrow \H_m \:,
\eeq
where~$k_m$ is the {\em{causal fundamental solution}} defined as the difference of the
advanced and retarded Green's operators,
\[ k_m := \frac{1}{2 \pi i} \left( s_m^\vee - s_m^\wedge \right) \::\: C^\infty_0(\scrM, S\scrM) \rightarrow 
\H_m^\infty \:. \]
The fermionic projector~$P$ can be written as an
integral operator involving a uniquely determined distributional
{\em{kernel}}~${\mathcal{P}} \in \D'(\scrM \times \scrM)$, i.e.\ (for details see~\cite[Section~3.5]{finite})
\beq \label{Pkerndef}
\bra \phi | P \psi \ket = {\mathcal{P}} \big( \overline{\phi} \otimes \psi \big)
\qquad \text{for all~$\phi, \psi \in C^\infty_0(\scrM, S\scrM)$}\:.
\eeq

A main application of our constructions is that the 
projection operator~$\chi_{(-\infty, 0)}(\Sig_m)$ gives rise to a distinguished
quasi-free ground state of the second-quantized Dirac field with the property that the
two-point distribution coincides with the kernel of the fermionic projector.
Indeed, applying Araki's construction in~\cite{araki1970quasifree}
gives the following result (see~\cite[Theorem~1.4]{hadamard}):
\begin{Thm} \label{thmstate}
There is an algebra of smeared fields generated by~$\Psi(g)$, $\Psi^*(f)$
together with a pure quasi-free state~$\omega$ with the following properties: \\[0.3em]
{\rm{(a)}} The canonical anti-commutation relations hold:
\[ \{\Psi(g),\Psi^*(f)\} = \bra g^* \,|\, k_m\, f \ket \:,\qquad
\{\Psi(g),\Psi(g')\} = 0 = \{\Psi^*(f),\Psi^*(f')\} \:. \]
{\rm{(b)}} The two-point distribution of the state is given by
\[ \omega \big( \Psi(g) \,\Psi^*(f) \big) = -\iint_{\scrM \times \scrM} g(p) \,{\mathcal{P}}(p,q) f(q) \: d\mu_\scrM(q)
\, d\mu_\scrM(y)\:. \]
\end{Thm} \noindent
The state~$\omega$ is referred to as the {\em{fermionic projector state}}
(or FP state)~\cite{fewster+lang}.

We finally note that, using the functional calculus, for
any non-negative bounded Borel function~$W$ we obtain in generalization of~\eqref{Pstar} the operator
\beq \label{PWdef}
P_W := -W(\Sig_m)\, k_m \::\: C^\infty_0(\scrM, S\scrM) \rightarrow \H_m \:,
\eeq
which can again be represented according to~\eqref{Pkerndef} as an integral operator
with a kernel~${\mathcal{P}}_W \in \D'(\scrM \times \scrM)$.
Again using Araki's construction (this time for the positive operator~$W(\Sig_m)$)
gives a corresponding quasi-free state.
In this way, the fermionic signature operator gives rise to a whole class
of distinguished quasi-free states, which we refer to as {\em{generalized fermionic projector states}}.
As shown in~\cite[Section~11]{rindler} in the example of two-dimensional Rindler space-time,
this makes it possible to obtain thermal states from the fermionic signature operator.

\section{Symmetries of the Fermionic Signature Operator} \label{secgensymm}
\subsection{Symmetries of Space-Time} \label{secsymmst}
Let~$\G$ be a Lie group (possibly non-compact, of finite dimension~$d \geq 0$,
where the case~$d=0$ is a discrete group).
In view of our applications, we want to allow for the possibility that~$\G$
acts only {\em{locally}} as a symmetry group. To this end, let~$\U \subset \G$
be an open neighborhood of the neutral element~$\e \in \G$
(by choosing~$\U=\G$, one recovers standard group actions).
To every group element~$\h \in \U$ we want to associate an isomorphism
of the spinor bundle~$S\scrM$. Moreover, these isomorphisms should be compatible
with the group operations, whenever the group multiplication stays in~$\U$.
In order to have the inverse element to our disposal, we assume that
the implication
\beq \label{Uinv}
\g \in \U \quad \Longrightarrow \quad \g^{-1} \in \U
\eeq
holds. This property can always be arranged by intersecting~$\U$
with the set~$\{\g^{-1} \:|\: \g \in \U\}$. This leads us to the following definition:

\begin{Def} \label{defPhi}
Let~$\U \subset \G$ be an open neighborhood of~$\e$ with the property~\eqref{Uinv}.
Moreover, let~$\Phi$ be a smooth mapping
\[ \Phi \:\in\: C^\infty\big(S\scrM \times \U, S\scrM \big) \]
with the following properties:
\begin{itemize}[leftmargin=2em]
\item[\rm{(i)}] $\Phi$ is compatible with the local group operations, i.e.\
\beq
\Phi_\g \circ \Phi_\h = \Phi_{\g \h} \qquad \text{for all~$\g, \h \in \U$ with~$\g \h \in \U$}\:, \label{compat1}
\eeq
where~$\Phi_\g := \Phi(.,\g)$. In view of~\eqref{Uinv}, this implies that~$\Phi_\g$ is a diffeomorphism on~$S\scrM$
and that~$\Phi_\e= \text{\rm{id}}_{S\scrM}$.
\item[\rm{(ii)}] $\Phi$ is compatible with the bundle projection~$\pi$, meaning that the following diagram commutes:
\beq \label{commutative}
\begin{array}{cccl}
\Phi \::\!\!\!&S\scrM \times \U & \longrightarrow & S\scrM \\
&\pi \Big\downarrow && \pi \Big\downarrow \\[0.5em]
&\scrM \times \U & \longrightarrow & \;\,\scrM \end{array}
\eeq
\item[\rm{(iii)}] The mapping~$f : \scrM \times \U \rightarrow \scrM$ defined by the lowest line of this
commutative diagram is a family of isometries, i.e.
\label{iii}
\beq \label{fdef}
(f_\h)^* g = g \qquad \text{for all~$\h \in \U\:.$}
\eeq
\item[\rm{(iv)}] $\Phi$ is compatible with Clifford multiplication and preserves the spin scalar product up to a sign, i.e.
\[ \gamma\big( (f_\h)_* u \big) = \Phi_\h \; \gamma(u) \: \Phi_\h^{-1} \qquad \text{and} \qquad
\Sl \Phi_\h \psi \,|\, \Phi_\h \phi \Sr_{f(p)} = \epsilon(\h)\: \Sl \psi | \phi \Sr_p \:, \]
valid for all~$\h \in \U$, $u \in T_p\scrM$ and~$\psi, \phi \in S_p\scrM$.
Here~$\epsilon(\h)$ is defined by
\[ \epsilon(\h) = \left\{ \begin{array}{c} 1 \\
-1 \end{array} \right\} \;\;\text{if~$f_\h$} \;\;
\left\{ \begin{array}{c} \text{preserves} \\
\text{reverses} \end{array} \right\} \;\;\text{the time orientation}\:. \]
\item[\rm{(v)}] $\Phi$ describes a symmetry of the external potential~$\B$, i.e.\
\[ \B\big( f_\h(p) \big) = \Phi_\h \; \B(p) \: \Phi_\h^{-1} \:. \]
\end{itemize}
We refer to~$\Phi$ as a {\bf{local group of isomorphisms of the spinor bundle}}~$S\scrM$.
\end{Def}

For clarity, we here explain our notation: Suppose that~$f_\h$ maps the space-time point~$p$ to~$q$.
Then the lower star is the usual derivative, i.e.\
\[ (f_\h)_*\big|_p = Df_h\big|_p \::\: T_p\scrM \rightarrow T_q\scrM \:. \]
The upper star denotes the pull-back defined by the identity
\[ \big((f_\h)^* g \big)_p(u,v) = g_q\big((f_\h)_* u, \,  (f_\h)_* v\big) \:. \]
We also note for clarity that, as a consequence of the commutativity of the diagram~\eqref{commutative},
also~$f$ is compatible with the group operations, i.e.\
\beq \label{compat2}
f_\g \circ f_\h = f_{\g \h} \qquad \text{for all~$\g, \h \in \U$ with~$\g \h \in \U$}\:.
\eeq
Again using~\eqref{Uinv}, this implies that~$f_\g$ is a diffeomorphism on~$\scrM$
and that~$f_\e= \text{id}_{\scrM}$.

\subsection{Unitary Symmetries on Hilbert Spaces of Dirac Solutions} \label{secUsymm}
Throughout this section, we fix a group element~$\h \in \U$.

\begin{Lemma} \label{lemmacauchy}
The diffeomorphism~$f_\h$ maps Cauchy surfaces to Cauchy surfaces.
\end{Lemma}
\Proof Recall that a Cauchy surface in a globally hyperbolic space-time is defined
to be a subset with the property that any non-extendible causal curve intersects the set exactly once
(see for example~\cite[p.~62]{beem}).
Being an isometry of~$\scrM$, the mapping~$f_\h$ clearly maps inextendible causal curves to
inextendible causal curves. Since~$f_\h$ is invertible, a curve~$\gamma$ intersects a subset~$\scrN \subset \scrM$
if and only if~$f_\h(\gamma)$ intersects~$f_\h(\scrN)$. This concludes the proof.
\QED

\begin{Lemma} \label{lemmaDiracpreserve}
Let~$\psi_m \in C^\infty(\scrM, S\scrM)$ be a solution of the Dirac equation.
Then the push-forward~$(\Phi_\h)_* \psi \in C^\infty(\scrM, S\scrM)$ defined by
\beq \label{Phipsi}
\big( (\Phi_\h)_* \psi \big)\big(f_\h(p) \big) := \Phi_\h\big(\psi(p)\big)
\eeq
is again a solution of the Dirac equation.
\end{Lemma}
\Proof It follows immediately from Definition~\ref{defPhi}~(iii)--(v) and~\eqref{Dirdef}
that the Dirac operator is invariant under the symmetry transformations.
Therefore, solutions are mapped to solutions.
\QED

\begin{Lemma} \label{lemmadense}
The push-forward~$(\Phi_\h)_*$ maps the following solution spaces to each other,
\[ (\Phi_\h)_* \::\: \H_m^\infty \rightarrow \H_m^\infty \:. \]
Moreover, this mapping is bijective and preserves the scalar product~$(.|.)_m$.
\end{Lemma}
\Proof Let~$\psi_m \in \H_m^\infty$.
In view of the unique solvability of the Cauchy problem, $\psi_m$ is determined by
its initial data~$\psi_m|_\scrN \in C^\infty_0(\scrN, S\scrM)$ on a Cauchy surface~$\scrN$.
By Lemma~\ref{lemmacauchy}, the transformed set~$f_\h(\scrN)$ is again a Cauchy surface.
Moreover, as the mapping~\eqref{Phipsi} maps compact sets to compact sets, 
the restriction of the transformed wave function~$(\Phi_\h)_* \psi_m$ to
the transformed Cauchy surface is again compact.
Moreover, since~$(\Phi_\h)_* \psi_m$ is again a solution of the Dirac equation 
(see Lemma~\ref{lemmaDiracpreserve}), it follows from the unique solvability of the
Cauchy problem for initial data on the transformed Cauchy surface
that~$(\Phi_\h)_* \psi_m$ again has spatially compact support.

In order to show that the scalar product is preserved, we make use of the fact that
it can be computed on any Cauchy surface. Hence
\begin{align*}
&(\psi_m | \phi_m)_m = \int_\scrN \Sl \psi_m \,|\, \gamma(\nu) \,\phi_m \Sr_p\: d\mu_\scrN(p) \\
&\;\; \overset{(*)}{=}\int_{f_\h(\scrN)} \Sl (\Phi_\h)_* \psi_m \,|\, \gamma(\nu) \,(\Phi_\h)_* \phi_m \Sr_q\: d\mu_{f_\h(\scrN)} (q) 
= \big( (\Phi_\h)_* \psi_m \big| (\Phi_\h)_* \phi_m \big)_m \:,
\end{align*}
where in~$(*)$ we used that~$\Phi_h$ preserves the Lorentzian metric and the
spin scalar product (see Definition~\ref{defPhi}~(iii) and~(iv)).
If~$f_\h$ reverses the time orientation, then both the spin scalar product
and the future-directed normal~$\nu$ change their signs, so that~$(*)$ again holds.
This completes the proof.
\QED
In view of this result, we can uniquely extend the operator~$(\Phi_h)_*$ by continuity to
a unitary mapping on~$\H_m$. 
For notational clarity, we denote this operator by~$U_m^\h$,
\beq \label{Uh}
U^\h_m := \overline{(\Phi_\h)_*} \::\: \H_m \rightarrow \H_m \qquad \text{unitary}\:.
\eeq

The above construction can immediately be extended to families of solutions,
simply by carrying it out pointwise for each~$m \in I$. This gives the following result:
\begin{Lemma} \label{lemmaunit}
The push-forward~$(\Phi_\h)_*$ maps the following 
spaces of families of solutions to each other,
\[ (\Phi_\h)_* \::\: \H^\infty \rightarrow \H^\infty \:. \]
Moreover, this mapping is bijective and preserves the scalar product~$(.|.)$.
It uniquely extends by continuity to a unitary operator on families of solutions,
\[ U^\h := \overline{(\Phi_\h)_*} \::\: \H \rightarrow \H \qquad \text{unitary}\:. \]
The operator~$U^\h$ acts pointwise in~$m$ and commutes with~$T$,
\beq \label{Uhpoint}
\big(U^\h \psi\big)_m = U^\h_m \,\psi_m \qquad \text{and} \qquad
T U^\h = U^\h T \:.
\eeq
\end{Lemma}

\subsection{Symmetries of the Fermionic Signature Operator} \label{secsymmsig}
\begin{Thm} \label{thmSfinite}
Assume that~$(\scrM, g)$ is weakly $m$-finite
(see Definition~\ref{defmfinite}). Then, up to a sign, the fermionic signature operator
is invariant under the symmetry transformation, meaning that
for every~$\h \in \U$,
\beq \label{USfinite}
(U^\h_m)^* \:\Sig_m\: U^\h_m = \epsilon(\h)\: \Sig_m \:.
\eeq
\end{Thm}
\Proof For any~$\psi_m, \phi_m \in \H_m$,
\begin{align*}
(& \psi_m \:|\: \Sig_m\, \phi_m)_m = \bra \psi_m | \phi_m \ket \\
&\overset{(*)}{=} \epsilon(\h)\:
\bra (\Phi_\h)_* \psi_m \,|\,  (\Phi_\h)_* \phi_m \ket
= \epsilon(\h)\: \big( (\Phi_\h)_* \psi_m \,|\, \Sig_m\, (\Phi_\h)_* \phi_m \big)_m \\
&\!\!\overset{\eqref{Uh}}{=} \epsilon(\h)\: \big( U^\h_m \psi_m \,|\, \Sig_m\, U^\h_m \phi_m \big)_m
= \epsilon(\h)\: \big( \psi_m \,|\, (U^\h_m)^* \,\Sig_m\, U^\h_m \phi_m \big)_m \:,
\end{align*}
where in~$(*)$ we used that~$\Phi_\h$ keeps the integration measure unchanged
and preserves the spin scalar product up to a sign.
\QED

\begin{Thm} \label{thmSinfinite}
Assume that~$(\scrM, g)$ satisfies the weak mass oscillation
property (see Definition~\ref{defweakMOP}). Then, up to a sign, the operator~$\Sig$
defined by~\eqref{Sigdef} is invariant under the symmetry transformation, i.e.
\beq \label{US}
(U^\h)^* \:\Sig\: U^\h = \epsilon(\h)\: \Sig \:.
\eeq
Moreover, the operator-valued measure~$dS_m$ in~\eqref{dSdef} is invariant up to a sign,
\beq \label{USweak}
(U^\h)^* \:d\Sig_m\: U^\h = \epsilon(\h)\: d\Sig_m \:.
\eeq

If~$(\scrM, g)$ satisfies the strong mass oscillation
property (see Definition~\ref{defstrongMOP}), then the fermionic signature operators
are all invariant up to a sign, i.e.
\beq \label{USstrong}
(U^\h_m)^* \:\Sig_m\: U^\h_m = \epsilon(\h)\:  \Sig_m \qquad \text{for all~$m \in I$} \:.
\eeq
\end{Thm}
\Proof For any~$\psi, \phi \in \H^\infty$,
\begin{align*}
(&\psi \:|\: \Sig\, \phi) = \bra \p \psi | \p \phi \ket \\
&= \epsilon(\h)\: \bra (\Phi_\h)_* \,\p \psi |  (\Phi_\h)_* \,\p \phi \ket
\overset{(*)}{=} \epsilon(\h)\: \bra \p \,(\Phi_\h)_* \psi \:|\:  \p\, (\Phi_\h)_* \phi \ket \\
&= \epsilon(\h)\: \big( (\Phi_\h)_* \psi \,|\, \Sig\, (\Phi_\h)_* \phi \big)
= \epsilon(\h)\: \big( U^\h \psi \,|\, \Sig\, U^\h \phi \big)
= \epsilon(\h)\: \big( \psi \,|\, (U^\h)^* \, \Sig\, U^\h \phi \big) \:,
\end{align*}
giving~\eqref{US}. Here the only major difference to the proof of Theorem~\ref{thmSfinite}
is that in~$(*)$ we used that the transformation~$\Phi_\h$ is independent of~$m$
and therefore trivially commutes with~$\p$.
The relation~\eqref{USweak} follows from the definition~\eqref{dSdef} and the fact
that~$U^\h$ commutes with~$T$.

If the strong mass oscillation property holds, \eqref{USstrong} follows from the computation
\[ \Sig_m \psi_m = \big( \Sig \psi \big)_m = \epsilon(\h)\: \big( (U^\h)^* \,\Sig\, U^\h \psi \big)_m
= \epsilon(\h)\: (U_m^\h)^* \,\Sig_m\, U_m^\h \psi_m \:, \]
where we used that both~$U^\h$ and~$\Sig$ act on~$\H$ pointwise for every~$m \in I$
(see~\eqref{Uhpoint} and~\eqref{Spoint}).
\QED

\subsection{Strongly Continuous Unitary Representations of the Symmetry Group}
Varying the group element~$\h$, we obtain a mapping
\beq \label{Umdef}
U_m \::\: \U \rightarrow \Lin(\H_m) \:,\qquad \h \mapsto U_m^\h
\eeq
(with~$U_m^\h$ as defined in~\eqref{Uh}).

\begin{Thm} \label{thmstrong}
The mapping~\eqref{Umdef}
is a local unitary representation of~$\G$ which is strongly continuous, i.e.
\[ \lim_{\h \rightarrow \g} \big\| U_m^\h \psi_m - U_m^\g \psi_m \big\|_m = 0
\qquad \text{for all~$\psi_m \in \H_m$}\:. \]
\end{Thm}
\Proof
From the compatibility with the group operations~\eqref{compat1} and~\eqref{compat2}
it is straightforward to verify that~$U_m$ is a local group representation.
Moreover, in view of~\eqref{Uh} this representation is obviously unitary.
Therefore, it remains to show strong continuity.
To this end, let~$\h(\tau)$ for~$\tau \in (-\delta, \delta)$ and~$\delta>0$
be a smooth curve in~$\U$ with~$\h(0)=\e$. Using the group properties, it suffices to show
strong continuity of~$U_m^{\h(\tau)}$ at~$\tau=0$.

\begin{figure}%
%
\psscalebox{1.0 1.0} 
{
\begin{pspicture}(-2,-1.6497028)(17.457415,1.6497028)
\definecolor{colour0}{rgb}{0.8,0.8,0.8}
\pspolygon[linecolor=colour0, linewidth=0.02, fillstyle=solid,fillcolor=colour0](0.91241515,1.6268288)(7.5724154,1.6318288)(7.4474154,1.4818288)(7.272415,1.2668289)(7.062415,1.0668288)(6.7774153,0.82182884)(6.582415,0.64182884)(6.372415,0.45682883)(6.1524153,0.24682884)(5.932415,0.026828842)(5.692415,-0.23317116)(5.512415,-0.44317114)(5.4474154,-0.54317117)(5.557415,-0.67817116)(5.732415,-0.89317113)(5.9074154,-1.0731711)(6.1324153,-1.2731712)(6.3024154,-1.4231712)(6.517415,-1.6331712)(2.4074152,-1.6231712)(2.5874152,-1.4381711)(2.822415,-1.2731712)(2.957415,-1.1331712)(3.0874152,-0.9681712)(3.1874151,-0.85817116)(3.3174152,-0.73817116)(3.1624153,-0.56817114)(3.007415,-0.38317117)(2.8574152,-0.18817116)(2.6574152,0.036828842)(2.457415,0.21182884)(2.2924151,0.33682883)(2.092415,0.48682883)(1.8974152,0.63182884)(1.6924151,0.80682886)(1.5324152,0.99682885)(1.3924152,1.1768289)(1.2224152,1.3768288)(1.0224152,1.5518289)
\rput[bl](4.6724153,-1.0231712){\normalsize{$K_0$}}
\psbezier[linecolor=black, linewidth=0.04](0.014081827,-1.5121711)(0.78549564,-1.3515044)(0.40598592,-1.423999)(1.6480713,-1.119837824503581)(2.8901567,-0.81567675)(4.3824983,-0.44179443)(5.6789775,-0.5665045)(6.9754567,-0.69121456)(7.122292,-1.0853434)(8.920749,-1.3731712)
\rput[bl](8.534081,-1.1448379){\normalsize{$\scrN$}}
\rput[bl](1.4874152,1.2418288){\normalsize{$\text{supp}\, \psi_m$}}
\rput[bl](7.6774154,0.7168288){\normalsize{$K_\tau := \supp \psi_m \cap \scrN_\tau$}}
\psbezier[linecolor=black, linewidth=0.08](3.2974153,-0.73817116)(4.046489,-0.6481712)(4.4928546,-0.5281712)(5.457415,-0.5481711578369141)
\psbezier[linecolor=black, linewidth=0.04](0.0040818276,-1.2221712)(0.77549565,-1.0615045)(1.4559859,-0.74899894)(2.6280713,-0.09483782450358148)(3.8001566,0.55932325)(5.1274986,0.61820555)(6.4239774,0.4934955)(7.7204566,0.36878544)(7.532292,-0.3603434)(8.890748,-0.40317115)
\rput[bl](8.224082,-0.044837825){\normalsize{$\scrN_\tau$}}
\psbezier[linecolor=black, linewidth=0.08](2.7224152,-0.043171156)(3.4574447,0.3595209)(4.1742854,0.45820656)(4.502415,0.5118288421630859)(4.830545,0.5654511)(5.667632,0.60341763)(6.452415,0.49682885)
\pscircle[linecolor=black, linewidth=0.02, fillstyle=solid,fillcolor=black, dimen=outer](3.8124151,-0.67067116){0.0975}
\pscircle[linecolor=black, linewidth=0.02, fillstyle=solid,fillcolor=black, dimen=outer](3.4074152,0.23932885){0.0975}
\rput[bl](3.4224153,-0.28317115){\normalsize{$q=f_{\h(\tau)}^{-1}(p)$}}
\rput[bl](3.737415,-1.0981711){\normalsize{$p$}}
\psbezier[linecolor=black, linewidth=0.02, arrowsize=0.05291667cm 2.0,arrowlength=1.4,arrowinset=0.0]{->}(7.537415,0.86182886)(6.639842,1.0172135)(6.4295397,1.3460596)(5.722415,0.6668288421630859)
\psbezier[linecolor=black, linewidth=0.02](3.3024151,-0.73817116)(3.038021,-0.48048308)(2.8421843,-0.14916174)(2.612415,0.06682884216308593)(2.382646,0.28281942)(2.033021,0.4995169)(1.7774152,0.7168288)(1.5218092,0.9341408)(1.3876461,1.2578194)(0.89741516,1.6418289)
\psbezier[linecolor=black, linewidth=0.02](3.2874153,-0.75817114)(3.136184,-0.8518092)(2.994806,-1.0824881)(2.8572755,-1.2202498095223069)(2.719745,-1.3580115)(2.5841439,-1.3651252)(2.4024153,-1.6331712)
\psbezier[linecolor=black, linewidth=0.02](5.4474154,-0.54817116)(5.611543,-0.32089317)(5.795036,-0.14337805)(5.937415,0.016828842163085937)(6.0797944,0.17703573)(6.646543,0.6941068)(6.817415,0.8418288)(6.9882874,0.9895509)(7.299794,1.2220358)(7.577415,1.6418289)
\psbezier[linecolor=black, linewidth=0.02](5.462415,-0.56317115)(5.600248,-0.6800962)(5.7335896,-0.90279835)(5.937415,-1.093171157836914)(6.1412406,-1.283544)(6.323209,-1.4002558)(6.5324154,-1.6431712)
\psbezier[linecolor=black, linewidth=0.02, arrowsize=0.05291667cm 2.0,arrowlength=1.4,arrowinset=0.0]{->}(7.827415,-0.25817117)(7.857415,-0.47317114)(7.817415,-0.68817115)(7.6724153,-0.9581711578369141)
\psbezier[linecolor=black, linewidth=0.02, arrowsize=0.05291667cm 2.0,arrowlength=1.4,arrowinset=0.0]{->}(6.6524153,-0.5881712)(6.857415,-0.35817116)(6.917415,-0.07817116)(6.932415,0.3018288421630859)
\rput[bl](7.897415,-0.79317117){\normalsize{$f_\tau$}}
\rput[bl](6.952415,-0.51317114){\normalsize{$f_\tau^{-1}$}}
\end{pspicture}
}
\caption{Strong continuity for wave functions with spacelike compact support.}%
\label{figcauchy}%
\end{figure}
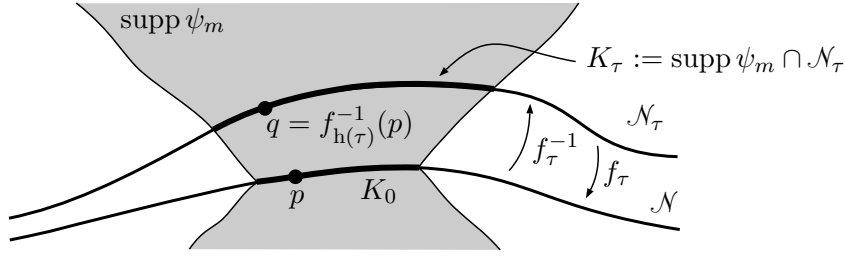

Our first step is to prove this strong continuity for smooth and spatially compact solutions, i.e.
\beq \label{step1}
\lim_{\tau \rightarrow 0} \big\|U_m^{\h(\tau)} \psi_m - \psi_m \big\|_m = 0 
\qquad \text{for all~$\psi_m \in \H_m^\infty$}\:.
\eeq
To this end, given~$\psi_m \in \H_m^\infty$ and a Cauchy surface~$\scrN$,
we set
\[ \scrN_\tau = f_{\h(\tau)}^{-1} \scrN \:. \]
Moreover, we define the compact sets
\[ K_\tau = \supp \psi_m \cap \scrN_\tau \:. \]
For a point~$p \in \scrN$ we denote the corresponding point on~$\scrN_\tau$
by~$q=f_{\h(\tau)}^{-1} p$. These notions are illustrated in Figure~\ref{figcauchy}.
According to~\eqref{Uh} and~\eqref{Phipsi},
\beq \label{Uformula}
\big( U_m^{\h(\tau)} \psi_m - \psi_m \big)(p) = \big( (\Phi_{\h(\tau)})_* \psi_m - \psi_m \big)(p)
=  \Phi_{\h(\tau)} \big( \psi_m(q) \big) - \psi_m(p) \:.
\eeq
Integrating over the Cauchy surface~$\scrN$, we obtain for the
norm corresponding to the scalar product~\eqref{print}
\begin{align*}
\frac{1}{4 \pi^2}\: \big\| U_m^{\h(\tau)} \psi_m - \psi_m \big\|_m^2
&= \int_{\scrN} \Big\| \Phi_{\h(\tau)} \Big( \psi_m \big( f_\tau^{-1}(p) \big) \Big)- \psi_m(p) \Big\|_p^2\,
d\mu_{\scrN}(p) \:,
\end{align*}
where the norm in the integrand denotes the pointwise norm on spinors
\[ \| \psi \|_p = \big( \Sl \psi \,|\, \gamma(\nu)\, \psi \Sr_p \big)^\frac{1}{2} \:. \]
Since all mappings as well as the wave function~$\psi_m$ are smooth, the integrand obviously tends to zero pointwise
as~$\tau \rightarrow 0$. Moreover, the integrand is supported in the set
\[ K_0 \cup f_\tau \big( K_\tau \big) \;\subset\; \scrN \:, \]
and it is bounded pointwise by
\[ \sup_{p \in \scrN, \;q \in \scrN_\tau} \Big( \|\psi_m(p)\|_p + \|\psi_m(q)\|_q \Big)^2 \:. \]
Using that all transformations are smooth and that the speed of propagation is finite,
the integrand is bounded and has
compact support, both uniformly for~$\tau \in [-\delta/2, \delta/2]$.
Therefore, we may take the limit~$\tau \rightarrow 0$ with the help of
Lebesgue's dominated convergence theorem to obtain~\eqref{step1}.

In order to extend the strong continuity to all of~$\H_m$, we use a
standard $3\varepsilon$-argument:
Given~$\varepsilon>0$ and~$\phi_m \in \H_m$, we choose~$\psi_m \in \H_m^\infty$
with~$\|\psi_m - \phi_m\|_m < \varepsilon$. Then
\begin{align*}
\big\| U_m^{\h(\tau)} \phi_m - \phi_m \big\|_m
&\leq \big\| U_m^{\h(\tau)} \big( \phi_m - \psi_m \big) \big\|_m
+ \big\| U_m^{\h(\tau)} \psi_m - \psi_m \big\|_m +
\big\| \psi_m - \phi_m \big\|_m \\
&< 2 \varepsilon + \big\| U_m^{\h(\tau)} \psi_m - \psi_m \big\|_m
\rightarrow 2 \varepsilon \:,
\end{align*}
where we used that~$U_m^{\h(\tau)}$ is unitary and
took the limit~$\tau \rightarrow 0$ using~\eqref{step1}.
Since~$\varepsilon$ is arbitrary, the result follows.
\QED

\subsection{Lie Algebra Representations and Commutators}
We denote the Lie algebra corresponding to~$\G$ by~${\mathfrak{g}} = T_\e \G$.
Given an element~$x \in {\mathfrak{g}}$ the exponential map gives a one-parameter
subgroup of~$\G$, which we denote by
\beq \label{hxtau}
h_x(\tau) := \exp_\e(\tau x) \:.
\eeq
\begin{Thm} \label{thmalgebra}
The following statements hold:
\begin{itemize}[leftmargin=2em]
\item[\rm{(i)}] For any~$x \in {\mathfrak{g}}$ and~$\psi_m \in \H_m^\infty$,
the derivative
\beq \label{Xdiff}
X \psi_m := -i \frac{d}{d\tau} \Big( U_m^{h_x(\tau)} \psi_m \Big) \Big|_{\tau=0}
\eeq
exists in~$\H_m$ and defines a linear operator
\beq \label{Xdense}
X \::\: \H_m^\infty \rightarrow \H_m^\infty \:.
\eeq
Considered as a densely defined operator on~$\H_m$ with domain~$\H_m^\infty$,
this operator is essentially self-adjoint.
\item[\rm{(ii)}] The mapping~$x \mapsto X$ is a representation of the Lie algebra~${\mathfrak{g}}$
on~$\H_m^\infty$.
\item[\rm{(iii)}] The following weak commutation relations hold,
\beq \label{weakcommute}
( X \psi_m \,|\, \Sig_m \phi_m)_m = ( \Sig_m \psi_m \,|\, X \phi_m)_m \qquad
\text{for all~$\psi_m, \phi_m \in \H_m^\infty$}\:.
\eeq
Moreover, if~$\Sig_m$ has a self-adjoint extension,
then the operators~$\overline{X}$ and~$\Sig_m$ commute in the sense that
their spectral measures commute (see~\cite[p.~271]{reed+simon}).
\item[\rm{(iv)}] In the setting of the weak mass oscillation property, the
spectral measure~$dE_{\rho,m}$ in~\eqref{dSdef}
commutes with the spectral measure of the operator~$\overline{X}$
(where~$X$ acts on~$\H^\infty$ pointwise in the mass, i.e.\ $(X \psi)_m := X\, \psi_m$).
Moreover, the following weak commutation relations hold,
\[ ( X \psi \,|\, \Sig \phi) = ( \Sig \psi \,|\, X \phi) \qquad
\text{for all~$\psi, \phi \in \H^\infty$} \:. \]
Finally, if~$\Sig$ has a self-adjoint extension,
then the operators~$\overline{X}$ and~$\Sig$ commute in the sense that
their spectral measures commute.
\end{itemize}
\end{Thm} \noindent
For clarity, we note that the assumption in~(iii) that~$\Sig_m$ must have a self-adjoint extension
is obviously satisfied if~$\Sig_m$ is a bounded operator. This is the case
if either space-time is $m$-finite (see~\eqref{stbound}) or if
the strong mass oscillation property holds (see Definition~\ref{defstrongMOP}).
If space-time is only weakly $m$-finite (see Definition~\ref{defmfinite}), the resulting
fermionic signature operator~$\Sig_m$ is only symmetric.
In this case, in order to obtain strong commutation relations, one must first
construct a self-adjoint extension of~$\Sig_m$
(as is done in Rindler space-time in~\cite[Section~8]{rindler}).
Similarly, if only the weak mass oscillation property holds, the operator~$\Sig$
defined by~\eqref{Sigdef} is only symmetric.

Before entering the proof, we remark that, knowing from Theorem~\ref{thmstrong}
that~$U_{h_x(\tau)}$ is a strongly continuous one-parameter group,
Stone's theorem (see for example~\cite[Theorem~VIII.8]{reed+simon}) implies that there is a self-adjoint generator, i.e.
\[ U_{h_x(\tau)} = e^{i \tau X} \qquad \text{with} \qquad
X : \D(\H_m) \subset \H_m \rightarrow \H_m \quad \text{self-adjoint}\:. \]
However, this abstract result does give explicit information on the domain.
In particular, Stone's theorem does not yield
that the domain contains the subset~$\H_m^\infty \subset \D(\H_m)$,
nor that this subspace is mapped to itself.
For this reason, we here prefer to use Chernoff's method~\cite{chernoff73}.

\begin{Lemma} For any~$x \in {\mathfrak{g}}$, the derivative~\eqref{Xdiff}
exists and is in~$\H^\infty_m$.
\end{Lemma}
\Proof
We fix~$x \in {\mathfrak{g}}$ and denote the one-parameter subgroup~\eqref{hxtau} 
for simplicity by~$h(\tau)$. Given~$\psi_m \in \H_m^\infty$, we know from~\eqref{Uformula} that
on a Cauchy surface~$\scrN$,
\[ \big( U_m^{\h(\tau)} \psi_m - \psi_m \big)(p) 
= \Phi_{\h(\tau)} \Big( \psi_m \big( f_\tau^{-1}(p) \big) \Big)- \psi_m(p) \:. \]
Since all transformations as well as the wave function~$\psi_m$ is smooth,
for any~$p \in \scrN$ the $\tau$-derivative exists,
\beq \label{diffpsi}
\frac{d}{d\tau} \big(U_m^{\h(\tau)} \psi_m\big)(p) =
\frac{d}{d\tau} \Phi_{\h(\tau)} \Big( \psi_m \big( f_\tau^{-1}(p) \big) \Big) \:.
\eeq
Moreover, these derivatives are bounded uniformly in~$\tau \in [-\delta/2, \delta/2]$,
locally uniformly in~$p \in \scrN$.
Exactly as in the proof of Theorem~\ref{thmstrong}, one sees that the support of
the function in~\eqref{diffpsi} is compact, again uniformly in~$\tau \in [-\delta/2, \delta/2]$.
Therefore, we may apply Lebesgue's dominated convergence theorem
to the difference quotient to conclude that the $\tau$-derivative exists in~$\H_m$.
Differentiating the Dirac equation
\[ (\Dir - m) \big(U_m^{\h(\tau)} \psi_m\big) = 0 \]
with respect to~$\tau$, we know furthermore that the derivative is again a solution.
Since the resulting derivative~\eqref{diffpsi} is obviously smooth and has
compact support on~$\scrN$, this solution is in~$\H^\infty_m$. This concludes the proof.
\QED

\begin{Lemma} For any~$x \in {\mathfrak{g}}$, the operator~$X$ in~\eqref{Xdense}
is essentially self-adjoint on~$\H_m$.
\end{Lemma}
\Proof From Lemma~\ref{lemmadense} we know that~$U_{h(\tau)}$ maps~$\H_m^\infty$
to itself. Moreover, using the group property,
\[ U_m^{\h(s)} \,U_m^{\h(\tau)} \psi_m = U_m^{\h(s+\tau)} \psi_m = U_m^{\h(\tau)} \,U_m^{\h(s)} \psi_m \:, \]
and differentiating the left and right with respect to $\tau$ at~$\tau=0$ gives
\[ U_m^{\h(s)} X \,\psi_m = X \,U_m^{\h(s)} \psi_m \:, \]
showing that the operators~$U_m^{\h(s)}$ and~$X$ commute on~$\H_m^\infty$.
Now we can apply the result by Chernoff~\cite[Lemma~2.1]{chernoff73} to conclude the proof.
\QED

\Proof[Proof of Theorem~\ref{thmalgebra}]
The previous two lemmas prove part~(i) of Theorem~\ref{thmalgebra}.
In order to prove~(ii), given~$x,y \in {\mathfrak{g}}$, the group properties imply that
for every~$\psi_m \in \H_m^\infty$ and all~$\tau \in \R$ with sufficiently small~$|\tau|$,
\[ \big(U_m^{\h_x(\tau)}\big)^{-1} \, \big(U_m^{\h_y(\tau)}\big)^{-1}\, U_m^{\h_x(\tau)} \, U_m^{\h_y(\tau)}
\,\psi_m = U_m^{\h_x(\tau)^{-1} \,\h_y(\tau)^{-1} \,\h_x(\tau) \,\h_y(\tau) } \,\psi_m \:. \]
In view of~\eqref{Xdiff} we may differentiate twice with respect to~$\tau$ at~$\tau=0$.
On the left side, this gives the commutator of operators~$-2 [X,Y]$.
On the right side, on the other hand, we may use the Lie algebra relation
\[ \h_x(\tau)^{-1} \,\h_y(\tau)^{-1} \,\h_x(\tau) \,\h_y(\tau) = \h_{[x,y]}(\tau^2) + \O\big(\tau^3 \big) \]
(which can be verified for example by using the Baker-Campbell-Hausdorff formula)
to obtain~$2 i Z$, where~$Z$ is the operator corresponding to the commutator~$z=[x,y] \in {\mathfrak{g}}$.
We thus obtain the relation~$Z = i [X,Y]$, proving~(ii).

For the proof of the weak commutation relations in~(iii),
we use that the operator~$\Sig_m$ with domain~$\H_m^\infty$ is symmetric
and that the operators~$U_m$ are unitary and commute with~$\Sig_m$. This gives
\[ \big( U_m^{\h_x(\tau)} \psi_m \,\big|\, \Sig_m \phi_m \big)_m = 
\big( \Sig_m \psi_m \,\big|\, \big(U_m^{\h_x(\tau)}\big)^{-1} \phi_m \big)_m \]
According to~\eqref{Xdiff} we may differentiate with respect to~$\tau$ at~$\tau=0$,
giving~\eqref{weakcommute}.

Now assume that~$\Sig_m$ has a self-adjoint extension (which we again
denote by~$\Sig_m$). According to~\eqref{USstrong}, we know that for all~$t \in \R$,
\[ (U^{\h_x(t)}_m)^* \:\Sig_m\: U^{\h_x(t)}_m = \Sig_m \:. \]
Using the spectral calculus, this equation also holds if~$\Sig_m$ is replaced
by powers of~$\Sig_m$ or by~$W(\Sig_m)$, where~$W$ is any bounded
Borel function. In particular, it follows that for all~$s,t \in \R$,
\[ e^{i s \Sig_m}\: U^{\h_x(t)}_m = U^{\h_x(t)}_m e^{i s \Sig_m} \:. \]
Noting that~$U^{\h_x(t)}_m = e^{i t \bar{X}}$, we can apply~\cite[Theorem~VIII.13~(c)]{reed+simon}
to conclude that~$\bar{X}$ and~$\Sig_m$ commute.
This concludes the proof of~(iii).

For the proof of~(iv), we make use of the fact that the group~$\G$ as
well as its Lie algebra act pointwise in~$m$.
Hence their representations on~$\H$ commute with~$T$.
Therefore, one can adapt the proof of~(iii) in a straightforward way
by inserting integrals over~$m$ to obtain the result.
\QED

\section{Symmetries of Generalized Fermionic Projector States} \label{secFP}
We now make precise in which sense the generalized fermionic projector
state preserves symmetries:
\begin{Thm} \label{thmgenFP}
For any non-negative bounded Borel function~$W$, the operator~$P_W$
defined in~\eqref{PWdef} has the symmetry property
\beq \label{PWtrans}
(U_m^\h)^* \,P_W \big( (\Phi_\h)_*\psi \big)  = 
\epsilon(\h)\: P_{W^\h}(\psi) \qquad \text{for all~$\psi \in C^\infty_0(\scrM, S\scrM)$}\:,
\eeq
where~$W^\h$ is defined by
\[ W^\h(\lambda) = W\big( \epsilon(\h)\: \lambda \big)\:. \]
Likewise, the kernel~${\mathcal{P}}_W \in \D'(\scrM \times \scrM)$ 
defined by~\eqref{Pkerndef} has the symmetry property
\beq \label{PWkernel}
\bra  (\Phi_\h)_* \phi \,|\, P_W (\Phi_\h)_* \psi \ket = 
\bra \phi \,|\, P_{W^h} \psi \ket
\qquad \text{for all~$\phi, \psi \in C^\infty_0(\scrM, S\scrM)$}\:.
\eeq
\end{Thm} \noindent
Before giving the proof of this theorem, we emphasize
a special case relevant for the applications:
\begin{Corollary} \label{corolltp}
Assume that~$f$ preserves the time orientation.
Then~$P_W$ is invariant under the symmetries in the sense
that for all~$\phi, \psi \in C^\infty_0(\scrM, S\scrM)$,
\begin{align}
(U_m^\h)^* \,P_W \big( (\Phi_\h)_*\psi \big)  &= P_W(\psi) \label{careful} \\
\bra  (\Phi_\h)_* \phi \,|\, P_W (\Phi_\h)_* \psi \ket &= 
\bra \phi \,|\, P_W \psi \ket \:.
\end{align}
\end{Corollary}

\Proof[Proof of Theorem~\ref{thmgenFP}]
We first derive the symmetries of the causal Green's operators~$s_m^\vee$
and~$s_m^\wedge$ (for basic definitions see for example~\cite[Section~2]{finite}).
Since~$\Phi_\h$ preserves the Dirac equation,
from the defining equation of the Green's operator we have for any~$\psi \in C^\infty_0(\scrM, S\scrM)$,
\[ \Phi_\h^{-1} \psi = \Phi_\h^{-1} (\Dir - m)\, s_m\, \psi = 
(\Dir - m)\, \Phi_\h^{-1} s_m\, \psi =
(\Dir - m)\, \big( \Phi_\h^{-1} s_m \Phi_\h \big) \, \Phi_\h^{-1} \psi \:. \]
Using that~$\Phi_h$ preserves the Lorentzian structure up to the time orientation, we obtain
\[ \Phi_\h^{-1} \,s^\wedge_m\, \Phi_\h = \left\{ \begin{array}{cl}
s_m^\wedge & \text{if~$f_\h$ preserves the time orientation} \\[0.3em]
s_m^\vee & \text{if~$f_\h$ reverses the time orientation} \:. \end{array} \right. \]
Being defined as the difference of the causal Green's operators
(see again~\cite[Section~2]{finite}), the causal fundamental solution~$k_m$ transforms according to
\[ \Phi_\h^{-1} \,k_m\, \Phi_\h = \epsilon(\h)\: k_m \:. \]
Since~$k_m$ maps to solutions, we can also write this identity as
\beq \label{kmtrans}
(U_m^\h)^* \,k_m \big( (\Phi_\h)_*\psi \big)  = \epsilon(\h)\: k_m(\psi) \:.
\eeq

Next, applying the spectral calculus to the symmetry statement in~\eqref{USfinite}
and~\eqref{USstrong}, we obtain
\[ W\Big((U^\h_m)^* \:\Sig_m\: U^\h_m \Big)
= (U^\h_m)^* \:W\Big(\epsilon(\h)\: \Sig_m\Big) \:U^\h_m = 
(U^\h_m)^* \:W^\h \big(\Sig_m\big) \:U^\h_m \:. \]
Multiplying by~\eqref{kmtrans} gives~\eqref{PWtrans}.

In order to prove~\eqref{PWkernel}, we use the
identity~$\bra \phi | \psi_m \ket = (k_m \phi \,|\, \psi_m)_m$, valid
for all~$\psi_m \in \H_m$ and~$\phi \in C^\infty_0(\scrM, S\scrM)$
(see~\cite[Proposition~3.1]{finite}).
We thus obtain
\begin{align*}
\bra \phi | P_{W^h} \psi \ket &= (k_m \phi \,|\, P_{W^h} \psi)_m \\
&\!\!\overset{\eqref{PWtrans}}{=} \epsilon(\h)\: (k_m \phi \,|\, (U_m^\h)^* \,P_W (\Phi_\h)_* \psi)_m
= \epsilon(\h)\: \big( U_m^\h \,k_m \phi \,\big|\, P_W (\Phi_\h)_* \psi \big)_m \\
&\!\!\overset{\eqref{kmtrans}}{=} \big(k_m \, (\Phi_\h)_* \phi \,\big|\, P_W (\Phi_\h)_* \psi \big)_m 
= \bra  (\Phi_\h)_* \phi \,|\, P_W (\Phi_\h)_* \psi \ket \:,
\end{align*}
giving the result.
\QED

We finally state an infinitesimal version of the above theorem. To this end,
for any~$x \in {\mathfrak{g}}$ we consider the curve~$h_x(\tau)$ in~\eqref{hxtau}
and introduce the Lie-type derivative
\begin{align*}
&L_x \::\: C^\infty_0(\scrM, S\scrM) \rightarrow C^\infty_0(\scrM, S\scrM) \:,\\
&\big(L_x \psi\big)(p) := \frac{d}{d\tau} \Big( \big(\Phi_{h_x(\tau)}\big)_* \psi \Big)(p) \Big|_{\tau=0}
\overset{\eqref{Phipsi}}{=} \frac{d}{d\tau} \Phi_{h_x(\tau)} \Big( \psi \big( f_{h_x(\tau)}^{-1}(p) \big) \Big) \Big|_{\tau=0} \:.
\end{align*}
\begin{Corollary} \label{corinf}
For any~$x \in {\mathfrak{g}}$, the generalized fermionic projector
state has the infinitesimal symmetries
\begin{align*}
\big(i X \phi_m \,\big|\, P_W \,\psi \big)_m +
\big(\phi_m \,\big|\, P_W  \,L_x\, \psi \big)_m &= 0 \\
\bra  L_x \eta \,|\, P_W \,\psi \ket + \bra \eta \,|\, P_W \,L_x\, \psi \ket &= 0\:,
\end{align*}
valid for all~$\eta, \psi \in C^\infty_0(\scrM, S\scrM)$ and~$\phi_m \in \H_m^\infty$.
\end{Corollary}
\Proof Follows immediately by evaluating the identities of Corollary~\ref{corolltp}
for~$\h=\h_x(\tau)$ and differentiating with respect to $\tau$ at~$\tau=0$.
Before differentiating~\eqref{careful}, one must take the inner product with~$\psi_m$,
making it possible to apply~\eqref{Xdiff}.
\QED

\section{Applications} \label{secappl}
We now consider two typical applications: infinitesimal symmetries
as described by Killing fields and discrete symmetries.
\subsection{Killing Symmetries} \label{seckilling}
In many applications, the symmetries of space-time are expressed in terms
of Killing fields. Since the commutator of two Killing fields is again Killing,
we may assume without loss of generality that the Killing fields form a Lie algebra~${\mathfrak{g}}$
of dimension~$d$,
\[ {\mathfrak{g}} \subset C^\infty(\scrM, T\scrM) \qquad \text{with} \qquad \dim {\mathfrak{g}}=d \geq 1\:. \]

In general, a Killing field does
{\em{not}} give rise to a corresponding symmetry of the fermionic signature operator,
as we now illustrate.

\begin{Example} {\bf{(The Minkowski drum)}} {\em{
Let~$\scrM \subset \R^{1,1}$ be a globally hyperbolic subset of two-dimensional
Minkowski space as considered in~\cite[Section~1.1]{drum}. Then the restriction
of the Killing fields of Minkowski space 
(the three generators of the Poincar{\'e} group in two dimensions)
are clearly Killing fields in~$\scrM$.
However, these Killing fields do not correspond to global symmetries of~$\scrM$.
Accordingly, the fermionic signature operator does {\em{not}} reflect the Killing symmetries.
This can be seen explicitly in the following counter example:

For the triangular domain
\[ \scrM = \big\{ (t,x) \in \R^{1,1} \:\big|\: 0 < t < \pi - |x| \big\} \:, \]
the fermionic signature operator is computed explicitly
in~\cite[Example~3.6]{drum}. Choosing the representation of the Dirac matrices
\[ \gamma^0 = \begin{pmatrix} 0 & 1 \\ 1 & 0 \end{pmatrix} \:,\qquad
\gamma^1 = \begin{pmatrix} 0 & 1 \\ -1 & 0 \end{pmatrix} \:, \]
the fermionic signature operator in the massless case~$m=0$ maps the plane wave solutions
\[ \psi^n_L(t,x) = \begin{pmatrix} 1 \\ 0 \end{pmatrix} e^{i n (x+t)} \:,\qquad
\psi^n_R(t,x) = \begin{pmatrix} 0 \\ 1 \end{pmatrix} e^{i n (x-t)} \]
with~$n \geq 1$ to each other. However, these plane waves are eigenfunctions of the Hamiltonian
\[ H = -i \gamma^0 \gamma^1 \partial_x = i \begin{pmatrix} 1 & 0 \\ 0 & -1 \end{pmatrix} \partial_x \]
with two different eigenvalues. Therefore, $H$ and~$\Sig_0$ do not commute.
Since~$H$ is the generator of time translations on~$\H_0$,
we conclude that the Killing field~$\partial_t$ does not correspond to a symmetry of~$\Sig_0$.
\QEDrem }}
\end{Example}

This example also explains why we need global symmetries as described by local group actions on~$\scrM$.
This leads us to impose an additional condition on the Killing fields:
\begin{Def} \label{defunifflow}
The Killing field~$K$ is {\bf{complete}} if there is an~$\varepsilon>0$
such that for every~$p \in \scrM$ the integral curve~$\gamma$ of~$K$ 
with~$\gamma(0)=p$ exists on~$(-\varepsilon, \varepsilon)$.
\end{Def} \noindent
By patching the solutions, this definition immediately implies that the integral curve exists on all of~$\R$.
We remark that in the special case
when the Killing field~$K$ describes the time translation symmetry of a static space-time,
the completeness of~$K$ follows from geodesic completeness (see~\cite[Theorem~2.1~(i)]{sanchez-static}).
We also remark that the Killing field is complete if it can be written as~$K=\frac{\partial}{\partial t}$
with a global time function~$t \in \R$.

Under the above assumption, the Killing symmetry can indeed be lifted to a local symmetry
of the spinor bundle:
\begin{Prp} \label{prplift} Assume that~${\mathfrak{g}}$ consists of complete Killing fields.
Then there is a Lie group~$\G$ with~$T_\e \G ={\mathfrak{g}}$
as well as a local group of isomorphism~$\Phi$ of~$S\scrM$
having all properties~{\rm{(i)--(v)}} in Definition~\ref{defPhi}. Moreover, the
corresponding space-time symmetry~$f$ in~\eqref{fdef}
is generated by the Killing fields in the sense that
\beq \label{Kback}
K(p) = \frac{d}{d\tau} f_{\exp_\e(\tau K)}(p)\big|_{\tau=0} 
\qquad \text{for all~$K \in {\mathfrak{g}}$}\:.
\eeq
\end{Prp} \noindent
For clarity, we remind the reader that the Lie algebra~${\mathfrak{g}}$
consists of vector fields on~$\scrM$ (see~\eqref{prplift}),
so that both sides of~\eqref{Kback} give a vector in~$T_p\scrM$.

\Proof[Proof of Proposition~\ref{prplift}]
Let~$\G$ be a Lie group with~$T_\e \G={\mathfrak{g}}$
(for example, one can choose the unique simply-connected Lie group with this property;
see~\cite[Theorem~3.15]{hall} or the ``converse of Lie's third theorem'' in~\cite[p.~108]{gilmore}).
Since the exponential map is locally invertible,
there is a neighborhood~$V$ of $0 \in {\mathfrak{g}}$ such that the restriction
\[ \exp_\e : V \rightarrow \exp_\e (V)=: \U \subset \G \]
is a diffeomorphism, and~$\U$ is an open neighborhood of~$\e$.
We denote its inverse by~$\log : \U \rightarrow V$.

Evaluating the integral curves of the Killing fields at~$\tau=1$, we obtain a smooth mapping
\[ E \::\: \scrM \times \mathfrak{g} \rightarrow \scrM \:, \]
which can be thought of as a realization of the exponential map on~$\scrM$.
Decomposing this mapping by the logarithm gives the desired
local group action on space-time
\beq \label{fhdef}
f \::\: \scrM \times \U  \rightarrow \scrM \:,\qquad
f(p,\h) := E\big(p, \log \h) \:.
\eeq
From the unique local characterization of Lie groups from their Lie algebras
(following from the convergence of the Baker-Campbell-Hausdorff formula,
see~\cite[Theorems~2.15.4 and~2.16.6]{varadarajan}; see also
the ``converse of Lie's second theorem'' in~\cite[p.~107]{gilmore}), this mapping is indeed
compatible with the local group operations~\eqref{compat2}
if the neighborhood~$\U$ is chosen sufficiently small.
Moreover, \eqref{Kback} follows immediately from~\eqref{fhdef}.

In order to construct the local group action on the spinor bundle,
we define~$\Phi_\h \psi(p)$
as the spinor obtained from~$\psi(p)$ by parallel transport
with respect to the spin connection along the integral curve of the Killing field~$\log \h$
through the point~$p$, evaluated at~$f_\h(p)$.
The compatibility with the group operation is verified as follows:
Let~$\psi, \phi \in S_p\scrM$ be two spinors. Then the corresponding
Dirac current is in the complexified tangent space,
\[ \Sl \psi | \gamma^j \phi \Sr_p\: \partial_j \big|_p =: v \in T_p^\C\scrM \:, \]
and since the spin connection is the lift of the Levi-Civita connection,
it follows that
\[ \Sl \Phi_\g \Phi_\h \psi | \gamma^j\, \Phi_\g \Phi_\h \phi \Sr_{f_{\g \h}(p)}
=  \Sl \Phi_{\g \h} \psi | \gamma^j\, \Phi_{\g \h} \phi \Sr_{f_{\g \h}(p)} =
\big((f_{\g \h})_* v\big)^j \:. \]
Moreover, knowing that the spinorial parallel
transport is unitary with respect to the spin scalar product
and that the last equation holds for all~$\psi, \phi \in S_p\scrM$, we conclude that
\[ \Phi^{-1}_\h \Phi_\g^{-1} \gamma^j\, \Phi_\g \Phi_\h = 
\Phi_{\g \h}^{-1} \gamma^j\, \Phi_{\g \h} \]
or, equivalently, that the following operators commute,
\[ \big[ A, \gamma^j \big] = 0 \qquad \text{with} \qquad A := \Phi_{\g \h} \Phi^{-1}_\h \Phi_\g^{-1} \:.\]
As a consequence, $A$ commutes with all elements of the Clifford algebra.
Since spinor representations are irreducible (see~\cite{lawson+michelsohn}),
it follows by Schur's lemma that~$A$
is plus or minus the identity. We conclude that
\beq \label{sign}
\Phi_\g \Phi_\h \big|_{S_p\scrM} = \pm \Phi_{\g \h} \big|_{S_p\scrM} \:.
\eeq
A continuity argument shows that, choosing the neighborhood~$\U$ sufficiently small,
only the plus sign appears. This concludes the proof.
\QED

We finally remark that by~\cite[Theorem~2.16.13]{varadarajan}, it is even possible
to define the action~$f$ on~$M$ on the whole group~$G$, if~$G$ is chosen to be
simply connected. However, in view of the sign ambiguity~\eqref{sign},
it is unclear to us how to construct the lift to~$S\scrM$ globally.
This is the reason why we restrict attention to local group actions.

\subsection{Discrete Symmetries} \label{secdiscrete}
As another typical application assume that~$\U=\G$ is a discrete group
acting as a group of isometries~$f : \scrM \times \G \rightarrow \scrM$ on space-time.
In order to get into the setting of Section~\ref{secgensymm}, 
the group of isometries on~$\scrM$ must be lifted to a group of isomorphisms of the spinor bundle~$S\scrM$.
More precisely, we need to construct a mapping~$\Phi$
which has all the properties in Definition~\ref{defPhi}.
This is a non-trivial task which involves a detailed knowledge of the group action~$f$
and the spin structure of~$S\scrM$.
Therefore, in order not to distract from the main topic of this paper,
we shall not enter this construction, but instead assume that~$\Phi$ is given
(however, in the simple example of Minkowski space and the group of parity and time reversal transformations,
the detailed construction is carried out in Section~\ref{exmink} below).
Then all the results of Sections~\ref{secUsymm}, \ref{secsymmsig} and~\ref{secFP} apply.
In particular, one obtains a unitary representation on~$\H_m$
(see~\eqref{Uh} and Lemma~\ref{lemmaunit}). Moreover, the fermionic signature
operator is invariant up to signs (see Theorems~\ref{thmSfinite} and~\ref{thmSinfinite}),
and the generalized fermionic projector state has the symmetry properties
stated in Theorem~\ref{thmgenFP}.

\section{Examples} \label{secex}
We now illustrate our constructions and results in a few examples in which
the fermionic signature operator has been studied previously.

\subsection{Minkowski Space} \label{exmink}
Let~$\scrM=\R^{1,3}$ be Minkowski space. We work with the Dirac equation in the
Dirac representation, i.e.\ $S\scrM = \scrM \times \C^4$ and
\[ \gamma^0 = \left( \begin{array}{cc} \1 & 0 \\ 0 & -\1 \end{array} \right) ,\qquad \gamma^\alpha = \left( \begin{array}{cc}
0 & \sigma^\alpha \\ -\sigma^\alpha & 0 \end{array} \right) \:, \]
where $\sigma^\alpha$ are the three Pauli matrices.
The Dirac equation reads
\[ \big( i \gamma^j \partial_j - m \big) \psi(x) = 0\:. \]

The symmetries of Minkowski space are described by the Poincar{\'e} group~${\mathcal{P}}$,
being the semi-direct product of translations with the Lorentz group,
\[ {\mathcal{P}} = \R^{1,3} \rtimes \mathrm{O}(1,3) \:. \]
The Lorentz group, in turn, is the semi-direct product of the
proper, orthochronous Lorentz group and a discrete group,
\[ \mathrm{O}(1,3) = \mathrm{SO}^+(1, 3) \rtimes \{ 1, P, T, PT \} \:, \]
where~$P$ is the parity transformation and~$T$ denotes time reflections, i.e.
\[ f_P(x) = (t, -\vec{x}) \qquad \text{and} \qquad
f_T\big(x) = (-t, \vec{x}) \:, \]
where~$x$ has the components~$(t,\vec{x})$.

We begin with the Killing symmetries. The Lie algebra of the Poincar{\'e} group
gives rise to ten Killing fields (4 translations, 3 rotations and 3 boosts).
The rotations and boosts are lifted infinitesimally by the well-known
transformations (see for example~\cite[Chapter~2]{bjorken})
\[ (\Phi_\h \psi)(\Lambda x) = S \psi(x) \]
with
\beq \label{inftrans}
dS = - \frac{i}{4}\:  d\Lambda_{jk} \:\sigma^{jk} \,S
\eeq
(where~$\Lambda \in \mathrm{SO}^+(1, 3)$ and~$\sigma^{jk} = i [\gamma^j, \gamma^k]/2$ are the bilinear covariants).
Integrating this equation globally gives rise to the usual representation
of the spin group on the spinor bundle.
With our concept of working with local group actions, one can avoid
topological issues and work with the 
proper, orthochronous Lorentz group instead of the spin group. Thus we choose
\[ \U \subsetneq \G := \R^{1,3} \rtimes \mathrm{SO}^+(1, 3) \:. \]
Restricting the group actions to a small subset of the identity
(necessarily with all rotation angles smaller than~$2\pi$),
and taking~$\Phi$ as the lift obtained by integrating~\eqref{inftrans},
we obtain a local group of isomorphism of the spinor bundle
(see Definition~\ref{defPhi}). Thus all results of Sections~\ref{secgensymm}
and~\ref{secFP} apply:
The unitary mappings~$\U_m^\h$ describe spatial translations and
time evolutions of Dirac solutions as well as rotations and boosts.
All these mappings are strongly continuous according to Theorem~\ref{thmstrong}.
Furthermore, Theorem~\ref{thmalgebra} shows that the corresponding generators~$X$
(being the momentum and angular momentum operators, the Hamiltonian
and infinitesimal boost operators) are all essentially self-adjoint.
Both the local representation of the group and the representation of the Lie algebra
on~$\H_m$ commute with the fermionic signature operator (see Theorems~\ref{thmSinfinite}
and~\ref{thmalgebra}).
Moreover, the resulting generalized fermionic projector states are
Lorentz invariant as specified in Corollaries~\ref{corolltp} and~\ref{corinf}.

We next consider the discrete group by choosing
\[ \U=\G = \{ 1, P, T, PT \} \:. \]
Setting\footnote{We remark that our lift of~$T$ does not agree with the
$T$-transformation in the physics literature
(see~\cite[Section~5.4]{bjorken} or~\cite[Section~3.6]{peskin+schroeder}),
where an {\em{anti}}-linear transformation is used.
Using the common notions in physics, our transformation corresponds to~$CT$,
where~$C$ is charge conjugation.}
\[ \big(\Phi_P \psi \big)\big(f_P(x) \big) = \gamma^0\,\psi(x) \qquad \text{and} \qquad
\big(\Phi_T \psi \big) \big(f_T(x) \big) = \gamma^5 \gamma^0\,\psi(x) \]
(where~$\gamma^5 = i \gamma^0 \gamma^1 \gamma^2 \gamma^3$),
the relations
\begin{align*}
\Sl \Phi_P \psi \,|\, \Phi_P \psi \Sr_{f_P(x)} &= \Sl \psi | \psi \Sr_x \\
\Sl \Phi_P \psi \,|\, \gamma^0 \,\Phi_P \psi \Sr_{f_P(x)} &= \Sl \psi \,|\, \gamma^0 \,\psi \Sr_x \\
\Sl \Phi_P \psi \,|\, \gamma^\alpha \,\Phi_P \psi \Sr_{f_P(x)} &= \Sl \psi \,|\, \big(-\gamma^\alpha \big) \,\psi \Sr_x \\
\Sl \Phi_T \psi \,|\, \Phi_T \psi \Sr_{f_T(x)} &= -\Sl \psi | \psi \Sr_x \\
\Sl \Phi_T \psi \,|\, \gamma^0 \,\Phi_T \psi \Sr_{f_T(x)} &= -\Sl \psi \,|\, \big(-\gamma^0 \big)\,\psi \Sr_x \\
\Sl \Phi_T \psi \,|\, \gamma^\alpha \,\Phi_T \psi \Sr_{f_T(x)} &= -\Sl \psi \,|\, \gamma^\alpha \,\psi \Sr_x
\end{align*}
(where~$\alpha=1,2,3$ denotes the spatial index) show that the transformation
indeed leaves Clifford multiplication invariant. Moreover, the $T$-transformation flips the sign of the
spin scalar product, in agreement with Definition~\ref{defPhi}~(iv).
Therefore, $\Phi$ is indeed a group of isomorphisms of the spinor bundle
(see Definition~\ref{defPhi}). Consequently, the results of Sections~\ref{secgensymm}
and~\ref{secFP} apply and show that the fermionic signature operator is
invariant under the group transformation, except that the $T$-transformation gives rise to a sign, i.e.\
\beq \label{Ttrans}
(U_m^T)^*\, \Sig_m\, U_m^T = -\Sig_m \:.
\eeq
These symmetry properties carry over to the generalized fermionic projector states
(see Theorem~\ref{thmgenFP}).

Taken together, these symmetry properties imply that all the positive-frequency
solutions (and similarly all the negative-frequency solutions) form an eigenspace of~$\Sig_m$.
Moreover, the $T$-transformation~\eqref{Ttrans} shows that the corresponding eigenvalues
have the same absolute value but opposite signs.
Hence these symmetry considerations determine the fermionic signature operator
(as computed in~\cite[Theorem~5.1]{infinite} to have eigenvalues~$\pm 1$)
up to a real prefactor.  The symmetry considerations also show that the generalized
fermionic projector states do not give more than the usual frequency splitting.

\subsection{Static and Ultrastatic Space-Times}
In {\em{static}} space-times, the symmetry under time translations is described by a timelike Killing field~$K=\partial_t$.
Integrating this Killing symmetry gives the group action~$f$ of~$\G=\R$ on~$\scrM$.
The lift~$\Phi$ to~$S\scrM$ is obtained simply by parallel transport of spinors along the Killing field.
Thus all results of Sections~\ref{secgensymm} and~\ref{secFP} again apply.
The resulting strongly continuous family of operators~$U_\h$ are the time evolution operators,
and its essentially self-adjoint generator~$X$ is the Dirac Hamiltonian.
Moreover, we conclude that the spectral measures of~$\overline{X}$ and~$\Sig_m$ commute.

In static space-times one also has the symmetry~$T$ under time reversals.
Similar as explained in Minkowski space, this again gives rise to the
symmetry~\eqref{Ttrans}.

For {\em{ultrastatic space-times}} it was shown in~\cite[Theorem~5.1]{infinite}
that the fermionic signature operator has eigenvalues $\pm 1$, and that
the corresponding eigenspaces coincide with the subspaces of positive and
negative frequency, respectively. This is consistent with the above results obtained
from the symmetry considerations, but of course the symmetry considerations
give much less information.

\subsection{Rindler Space-Time}
In~\cite{rindler} the fermionic signature operator is computed in two- and four-dimensional
Rindler space-time. We now discuss to which extent these results can be obtained from
symmetry considerations.

In two-dimensional Rindler space-time, there is a timelike Killing field~$K$ describing
Lorentz boosts. Therefore, space-time is static with respect to so-called Rindler time~$\tau$
(defined by the relation~$\partial_\tau =K$; for details see~\cite[Section~10]{rindler}),
and the symmetry considerations of the previous section imply that~$\Sig_m$ commutes with the
Hamiltonian in Rindler time and is anti-symmetric under reversals of
Rindler time~\eqref{Ttrans}.
In~\cite[Theorem~10.1]{rindler} it is shown that~$\Sig_m$ is indeed a multiple
of the Hamiltonian. This is compatible with the symmetry considerations but
clearly is a much stronger result.

In four-dimensional Rindler space-time, the two additional spatial coordinates~$y$ and~$z$
give rise to an additional symmetry group~$\R^2 \rtimes \mathrm{O}(2)$.
Thus the total symmetry group is
\beq \label{symmrind}
\G = \R^1 \times \big( \R^2 \rtimes \mathrm{O}(2) \big) \:.
\eeq
The additional symmetry means that, after separating the $y$- and~$z$-dependence by
plane waves with momenta~$k_y$ and~$k_z$, the fermionic signature operator must
depend only on~$k_y^2 + k_z^2$. This is compatible with the result in~\cite[Theorem~13.2]{rindler},
but again the explicit computation of the fermionic signature operator gives more
detailed information.

\subsection{Closed Friedmann-Robertson-Walker Space-Times}
In~\cite[Section~5]{finite} and \cite[Section~6]{infinite}
the fermionic signature operator is computed in spatially symmetric space-times.
The symmetry results in the present paper show that the fermionic signature operator
is diagonal on the spatial modes (i.e.\ the eigenspinors of the spatial Dirac operator;
for details see~\cite{moritz}). Thus our results give a more abstract explanation
for the separation procedure used in the the above-cited papers.

\subsection{A Plane Electromagnetic Wave}
In~\cite{planewave} the fermionic signature operator is computed in
Minkowski space in the presence of an electromagnetic potential
of the form of a plane wave, i.e.
\[ A = A(t+x) \:. \]
As shown in~\cite[Theorem~5.5]{rindler}, the fermionic signature operator
has eigenvalues~$\pm 1$, and the corresponding eigenspaces are
the solutions of positive and negative momentum~$u$ in the
separation ansatz in null direction
\[ \psi_m(t,x,y,z) = e^{-iu(t-x)} \: \chi_m(t+x,y,z) \:. \]

This result can again be partly understood from symmetry considerations:
The symmetries are again described by~\eqref{symmrind},
where the group~$\R^1$ now describes translations in the null direction~$(1,-1,0,0)$,
and the group~$\R^2 \rtimes \mathrm{O}(2)$ again acts on the
spatial coordinates~$y$ and~$z$. The momentum~$u$ is precisely the spectral
parameter of the generator~$X$ of the translations in null directions.

\Thanks {{\em{Acknowledgments:}}
We would like to thank Olaf M\"uller for helpful discussions. 
F.F.\ is grateful to the Instituto Superior T{\'e}cnico in Lisbon for hospitality and support.

\providecommand{\bysame}{\leavevmode\hbox to3em{\hrulefill}\thinspace}
\providecommand{\MR}{\relax\ifhmode\unskip\space\fi MR }
\providecommand{\MRhref}[2]{%
  \href{http://www.ams.org/mathscinet-getitem?mr=#1}{#2}
}
\providecommand{\href}[2]{#2}

\end{document}